\documentclass{revtex4}

\usepackage{amsmath,amssymb}
\usepackage{graphicx}
\usepackage{dcolumn}
\usepackage{bm}

\begin {document}

\title{Crossover Time in Relative Fluctuations Characterizes the
Longest Relaxation Time of Entangled Polymers}

\author{Takashi Uneyama}
\affiliation{%
  JST-CREST, Institute for Chemical Research, Kyoto University, Gokasho,
  Uji 611-0011, Japan
}%
\affiliation{%
  School of Natural System, College of Science and Engineering,
  Kanazawa University, Kakuma, Kanazawa 920-1192,
  Japan
}%

\author{Takuma Akimoto}
\affiliation{%
  Department of Mechanical Engineering, Keio University, Yokohama 223-8522, Japan
}%

\author{Tomoshige Miyaguchi}
\affiliation{%
Department of Mathematics Education, 
Naruto University of Education, Tokushima 772-8502, Japan}


\date{\today}

\begin{abstract}
  In entangled polymer systems, there are several
  characteristic time scales, such as the entanglement time
  and the disengagement time. In molecular simulations, the
  longest relaxation time (the disengagement time) can be determined by the
  mean square displacement (MSD) of a segment or by the shear relaxation
  modulus. {Here, we propose the relative fluctuation analysis method, which is originally
   developed for characterizing large fluctuations, to determine the longest
  relaxation time from the center of mass trajectories of polymer chains (the time-averaged MSDs).
  Applying the method to simulation data of
 entangled polymers (by the slip-spring model and the simple reptation model),
 we} provide a clear evidence that the
  longest relaxation time
  is estimated as the crossover time in the relative
  fluctuations.
\end{abstract}

\maketitle


\section{Introduction}
\label{introduction}

Polymer melts and solutions exhibit interesting dynamical behavior.  In
particular, if the degree of polymerization is large, characteristic
dynamical behavior due to the entanglement effect can be
observed\cite{Doi1986}.  For example, viscoelasticity with a very long
relaxation time (which strongly depends on the molecular weight)
and anomalous diffusion of a segment occur as a result of the
entanglement effect.  Dynamics of entangled polymers have been studied
extensively by various methods\cite{Doi1986,Watanabe1999,McLeish2002},
including theoretical modeling and simulations.

There are several characteristic time scales in entangled polymer systems,
such as the entanglement time $\tau_{e}$, the Rouse time $\tau_{R}$, and
the disengagement time (the longest relaxation time) $\tau_{d}$.  In each
characteristic time scale, mean square displacements (MSDs)
show different scalings.  For example, the Doi-Edwards tube model gives
the following time dependence of the MSD of a single segment\cite{Doi1986}:
\begin{equation}
 \label{tube_model_msd_segment}
 \langle [\bm{r}_{i}(t) - \bm{r}_{i}(0)]^{2}\rangle
  \propto
\begin{cases}
 t^{1/2} & (t \lesssim \tau_{e}) \\
 t^{1/4} & (\tau_{e} \lesssim t \lesssim \tau_{R}) \\
 t^{1/2} & (\tau_{R} \lesssim t \lesssim \tau_{d}) \\
 t & (\tau_{d} \lesssim t),
\end{cases}
\end{equation}
where $\bm{r}_{i}(t)$ is the position of the $i$-th segment
at time $t$ and $\langle \dots \rangle$ represents the
ensemble average.
Anomalous diffusion of a single segment in the Doi-Edwards
tube model originates from the following two mechanisms.  One is a
dynamical property of a single Rouse chain.
The MSD of a segment of a Rouse chain shows subdiffusive behavior, $\langle [\bm{r}_{i}(t) - \bm{r}_{i}(0)]^{2}\rangle
\propto t^{1/2}$, at the short time scale ($t \lesssim
\tau_{R}$)\cite{Doi1986}. 
The other is a constraint by a tube (which is formed by
  surrounding chains). Namely, each segment can move only along the tube,
  which has a fractal geometry similar to a random walk trajectory, thereby
  causing a subdiffusive transport characterized by the exponent $1/2$ for $\tau_{e} \lesssim t
  \lesssim \tau_{d}$. The subdiffusive exponent $1 / 4$ for $\tau_{e}
\lesssim t \lesssim \tau_{R}$ results from the combination of these two
mechanisms. Thus, using the MSD of a single segment, one can
estimate $\tau_{e}$, $\tau_{R}$ and
$\tau_{d}$\cite{Kremer1990,Putz2000,Sukumaran2009}.

It should be noticed that numerical prefactors in
Eq. \eqref{tube_model_msd_segment} (which are not explicitly shown) depend
on the index $i$  if the target segment is close to chain
ends\cite{Sukumaran2009}. (And if the strong $i$-dependence is
observed, validity of Eq. \eqref{tube_model_msd_segment} is no longer guaranteed.)
Although this does not cause serious problems for
relatively fine-scale models (such as the Kremer-Grest
model\cite{Kremer1990,Putz2000,Sukumaran2009}), it may be crucial for
highly coarse-grained models (such as the RaPiD (responsive particle
dynamics) model\cite{Kindt2007,Padding2011}). In some coarse-grained
models, resolutions of the models are not sufficient to resolve the MSDs of
segments.  Therefore, for such coarse-grained models, it is
physically reasonable to employ the MSD of the center of mass (CM) of a
chain, instead of that of a single segment.  In contrast to
the MSD of a single segment, the MSD of the CM is given by
\begin{equation}
 \label{tube_model_msd_cm}
 \langle [\bm{r}_{\text{CM}}(t) - \bm{r}_{\text{CM}}(0)]^{2}\rangle
  \propto
\begin{cases}
 t & (t \lesssim \tau_{e}) \\
 t^{1/2} & (\tau_{e} \lesssim t \lesssim \tau_{R}) \\
 t & (\tau_{R} \lesssim t),
\end{cases}
\end{equation}
where $\bm{r}_{\text{CM}}(t)$ is the position of the CM. We note that the
scaling exponent in the MSD of the CM does not change around $t \approx
\tau_{d}$.
Roughly speaking, this is because the constraint by a tube apparently
disappears by taking the average with respect to all segments.  It seems to
be difficult to extract an information of $\tau_{d}$ from the MSD of the CM
straightforwardly.

For highly coarse-grained models, thus, usually other
physical quantities such as the shear relaxation modulus and the end-to-end
vector relaxation function have been employed to determine
the long time relaxation behavior.
{ However, it is not always easy to determine the long time
relaxation behavior accurately. For example, to obtain the
accurate shear relaxation modulus data, we need to compute
the autocorrelation functions of stress tensors for large number of
statistically independent samples. Such calculation becomes demanding for
interacting many chain models. The end-to-end vector relaxation
function can be computed more accurately than the relaxation modulus,
but some highly coarse-grained models
such as the RaPiD model do not provide the information of end-to-end
vectors. Therefore, alternative methods to determine the long time
behavior will be useful for some simulation models, and thus are worth developing.
A simple but fascinating idea is to utilize some tools or methods developed in the fields other
than polymer physics to analyze simulation data of entangled polymers.

In this work, we employ a method which is originally
developed in L\'{e}vy statistics\cite{Bardou2002}, to analyze the long time
relaxation behavior of entangled polymers. Recently, various theoretical
tools have been developed for the continuous time random walk (CTRW)
model\cite{Metzler2000}, which is a typical trap model.
An interesting recent finding is that the distribution of the
time-average MSD (TAMSD) can be utilized to characterize some dynamical
properties\cite{He2008,Jeon2010,Akimoto2011,Miyaguchi2011,Miyaguchi2011a,ctrw-inpreparation}. The TAMSD for a position $\bm{r}$ is defined as 
\begin{equation}
 \label{tamsd}
\overline{\delta^2(\Delta ;t)}  \equiv \frac{1}{t-\Delta}\int_0^{t-\Delta} dt'
 \, [ \bm{r}(t'+\Delta)- \bm{r}(t')]^2 ,
\end{equation}
with $\Delta$ and $t$ being the time difference and the observation time.
The relative fluctuation (RF) of the TAMSD 
\begin{equation}
 \label{relative_fluctuation_tamsd}
R(t;\Delta)  \equiv \frac{\langle |\overline{\delta^2(\Delta ;t)} -
 \langle \overline{\delta^2(\Delta; t)} \rangle |\rangle}{\langle
 \overline{\delta^2(\Delta; t)} \rangle},
\end{equation}
is useful for characterizing non-Gaussian fluctuations and 
ergodicity breaking in diffusion processes\cite{Akimoto2011, He2008, Miyaguchi2011, Miyaguchi2011a}.
For the CTRW trajectories, by analyzing the $t$-dependence of the RF, we
can determine some important information of the 
trapping-time distribution, such as the power-law exponent
and relaxation time\cite{Miyaguchi2011a, Akimoto2011}.

Although the CTRW is not equivalent to models for
entangled polymers, the RFs in polymer systems may also characterize
some long time relaxation behavior.
If the RF analysis (RFA) works well for
entangled polymers, it will serve as a new and complementary tool to study the
long time relaxation behavior.
As far as the authors know, such an analysis has never been performed
for entangled polymer systems. In this work we perform the RFA for the
CM trajectory data of entangled polymer models and show that the
information of the long time relaxation behavior (such as the longest
relaxation time $\tau_{d}$) can be observed.
Therefore the RFA can be utilized as a new
and complementary method.
}

\section{Simulation Results}
\label{simulation_results}

In this section, we perform
simulations for entangled polymers and analyze MSDs of CMs.  So far,
various simulation models have been proposed and utilized to study
entangled polymers\cite{Kremer1990,Hua1998,Masubuchi2001,Doi2003,
Nair2006,Likhtman2005, Uneyama2011,Kindt2007,Padding2011}. In this work,
we employ two simple models among them.
One is the slip-spring model\cite{Likhtman2005,Uneyama2011} and the other
is the simple Doi-Edwards type reptation model\cite{Doi1978}.
We calculate the RFs of the TAMSDs, Eq.~\eqref{relative_fluctuation_tamsd},
for the CM trajectories ($\bm{r}_{\text{CM}}(t)$), and show that the RFs show
clear crossovers around $t \simeq \tau_{d}$.

\subsection{Slip-Spring Model}
\label{slip_spring_model}
First we perform simulations of the slip-spring
model\cite{Likhtman2005,Uneyama2011}. The slip-spring model is a variant of
slip-link models\cite{Hua1998,Masubuchi2001,Doi2003,Nair2006}, which
reproduce diffusive and rheological properties well. In the slip-spring
model, a polymer chain is described as an ideal chain and the entanglements
are mimicked by so-called slip-springs. One end of a slip-spring is
anchored in space while another end can slip along the chain. Moreover,
slip-springs are dynamically reconstructed at chain ends.
This model can reasonably reproduce linear and nonlinear rheological
properties, while it lacks some mechanisms such as the constraint
release (CR) or the convective constraint release (CCR)\cite{Watanabe1999,McLeish2002}.
We perform
simulations for different numbers of beads (polymerization indices)
$N$. Other parameters are set to be the standard values; the average number
of segments between slip-springs $N_{0} = 4$, the slip-spring strength
$N_{s} = 0.5$, the friction coefficient of a slip-spring $\zeta_{s} / \zeta
= 0.1$ (with $\zeta$ being the friction coefficient of a segment).
As elementary length and time scales of this model, we use
the segment size $b$ and the characteristic time of a segment $\tau_{0}
\equiv \zeta b^{2} / k_{B} T$. ($k_{B}$ and $T$ are the Boltzmann
constant and the temperature, respectively.)
For each
$N$ value, simulations are performed for $1024$ polymer chains sampled from
an equilibrium distribution. (Details of simulation model and algorithm
used in this work are described in Ref.\cite{Uneyama2011})

\begin{figure}[h]
 \includegraphics[width=1.0\linewidth,clip]{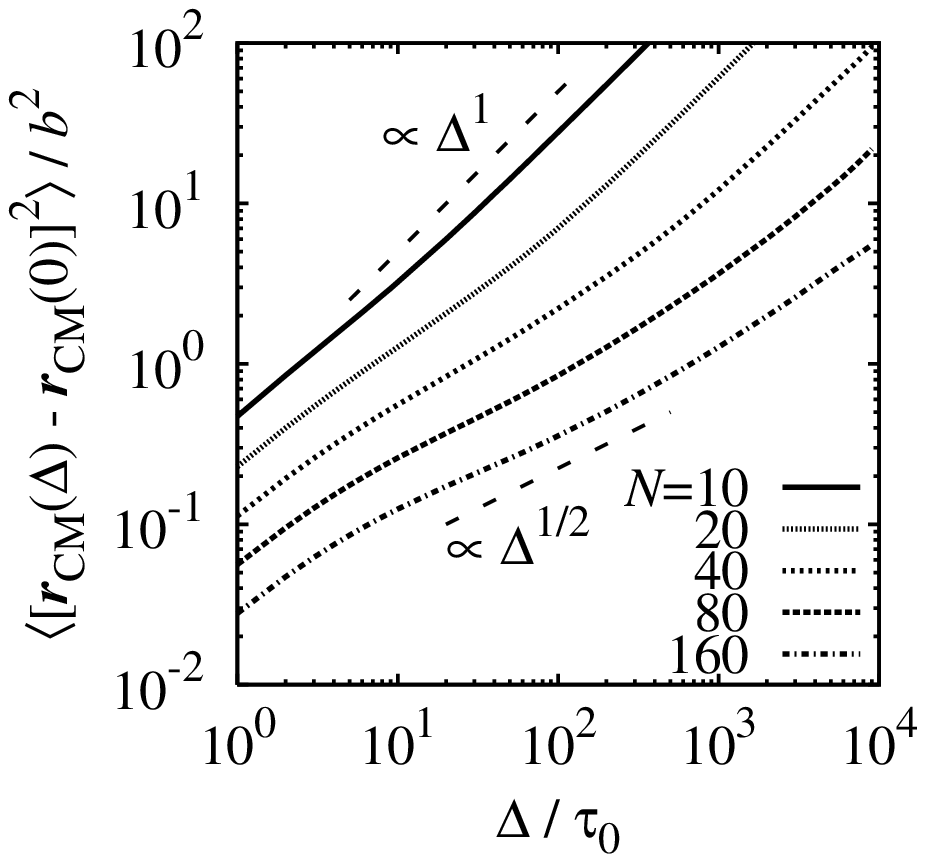}
\caption{\label{slip_spring_eamsd}}
\end{figure}

Figure~\ref{slip_spring_eamsd} shows the ensemble-averaged MSDs (EAMSDs) of
CMs for $N = 10, 20, 40, 80$, and $160$. (The CM is determined as the
average position of the beads which compose a polymer chain.)  For large
$N$ cases, the $t$ dependence of the MSDs is well described by
Eq.~\eqref{tube_model_msd_cm}.  Thus information of two
characteristic times, $\tau_{e}$ and $\tau_{R}$, can be obtained. However,
as we mentioned in Section \ref{introduction}, we cannot obtain any
information about $\tau_{d}$ from the EAMSDs for CMs.  Here
we perform the RFA of the TAMSDs for CMs.
{As will be shown in Appendix \ref{appendix_effect_of_time_interval_size_delta}, 
it is difficult to obtain the RF accurately if $\Delta$ is too small. Also, if $\Delta$ is too large, 
the RF shows only Gaussian fluctuation. Therefore, the time difference $\Delta$ should be chosen to be some moderate values. 
Here, we employ $\Delta/ \tau_0 =10$ for the current slip-spring simulations.}

\begin{figure}[p]
 \includegraphics[width=1.0\linewidth,clip]{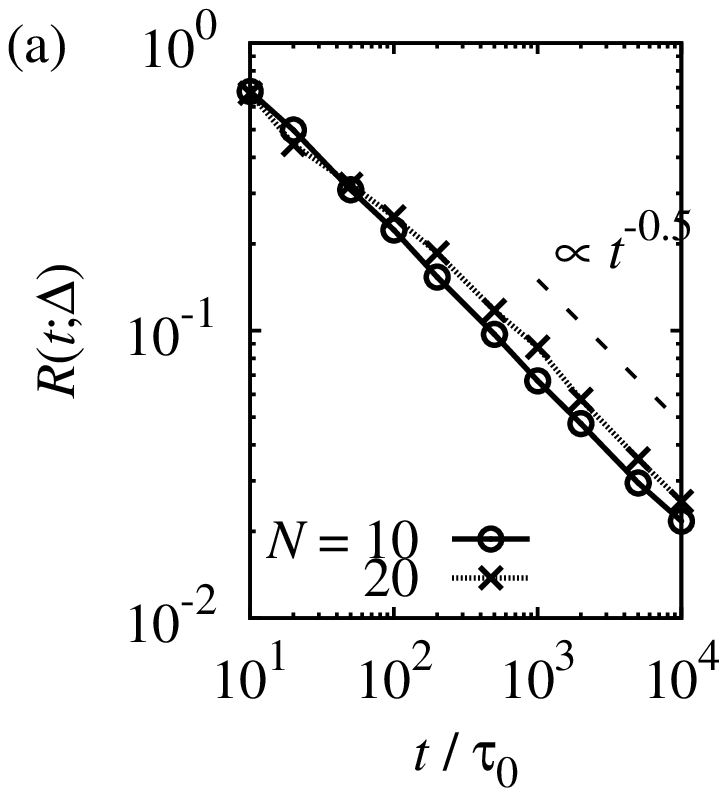}
 \includegraphics[width=1.0\linewidth,clip]{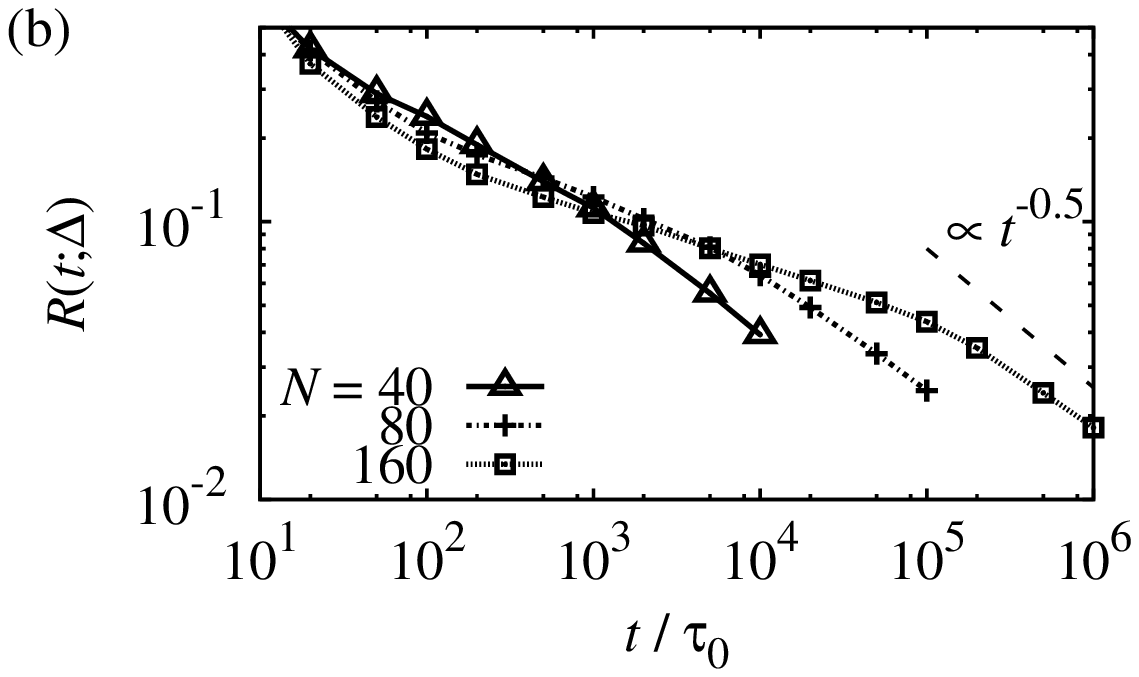}
\caption{\label{slip_spring_tamsd_fluctuation_delta10}}
\end{figure}

{Figure~\ref{slip_spring_tamsd_fluctuation_delta10} shows
the RF for different $N$.}
For large $N$ cases ($N \gtrsim 40$), the RF is
fitted well by the following form except for a very short time region:
\begin{equation}
 \label{relative_fluctuation_tamsd_cm}
 R(t;\Delta) \propto
  \begin{cases}
   t^{-\alpha} & (t \lesssim \tau_{c}) \\
   t^{-0.5} & (t \gtrsim \tau_{c})
  \end{cases},
\end{equation}
where $\alpha$ is a power-law exponent ($0 < \alpha < 0.5$
) and $\tau_{c}$ is the
crossover time.
The crossover time, $\tau_{c}$, is much longer than the
crossover time for the EAMSD  {and strongly depends on $N$}.
{We cannot determine the crossover time accurately for
small $N$ ($N \lesssim 20$,
Fig.~\ref{slip_spring_tamsd_fluctuation_delta10}(a)).
For $N \gtrsim 40$
(Fig.~\ref{slip_spring_tamsd_fluctuation_delta10}(b)), we can
determine the crossover time from the RF data easily. The crossover time $\tau_{c}$
strongly depends on $N$. (The power-law exponent $\alpha$ also depends
on $N$. Fitting gives $\alpha = 0.31, 0.24$ and $0.19$ for $N = 40, 80$ and $160$,
respectively. We will discuss this in Section~\ref{discussions}.)
} Such behavior of $\tau_{c}$ seems to be
consistent with the behavior of the longest relaxation time $\tau_{d}$ determined
from the shear relaxation modulus\cite{Uneyama2011}.
Actually, the $N$ dependence of $\tau_{c}$ obtained from the RFA
 is quite similar to that of $\tau_{d}$ (see Fig.~\ref{slip_spring_relaxation_and_crossover_times}). 
For $N \gtrsim
40$, $\tau_{d}$ and $\tau_{c}$ show the similar power-law type $N$
dependencies ($\tau_{d} \propto N^{3.48}$ and $\tau_{c} \propto
N^{3.51}$).

{For $N \lesssim 20$, the $N$ dependence of $\tau_{d}$ is close to one of
unentangled polymers ($\tau_{d} \propto N^{2}$) and in this region
$\tau_{c}$ is not determined accurately (see Fig.~\ref{slip_spring_tamsd_fluctuation_delta10}(a)).}
Possibly, this implies that we cannot extract the
information about the long time relaxation behavior from the RFA if
$\tau_{d}$ is comparable to or less than $\Delta$. Similar trend is
found in Appendix \ref{appendix_effect_of_time_interval_size_delta}.
Although behavior of the RF of the TAMSDs depends on parameters such as $N$
and $\Delta$, we have qualitatively the same result for other cases as long as $N$ is sufficiently large.
Thus, in the slip-spring model, we conclude that $\tau_{c}$
is essentially the same as $\tau_{d}$ apart from the numerical constant.

\subsection{Discrete Reptation Model}

To confirm whether the relation between the cutoff time and the longest
relaxation time is a general feature in entangled polymer systems, second
we perform simulations of the simple Doi-Edwards type reptation model. This
model is one of the simplest models for entangled polymers, yet describes
some essential features of entangled polymer dynamics qualitatively. The
details of the model and the numerical scheme used in the 
  simulations are described in Appendix
\ref{appendix_discrete_reptation_model}.  Although several important
mechanisms (such as the constraint
release)\cite{Doi1986,Watanabe1999,McLeish2002} are not considered in this
model, we believe that the incorporation of these mechanisms
do not change the RF behavior of the TAMSD qualitatively.  Therefore, if it
gives qualitatively the same result as the slip-spring model, we consider
that the RFA can be applied to most of models for entangled polymers.

\begin{figure}[h]
 \includegraphics[width=1.0\linewidth,clip]{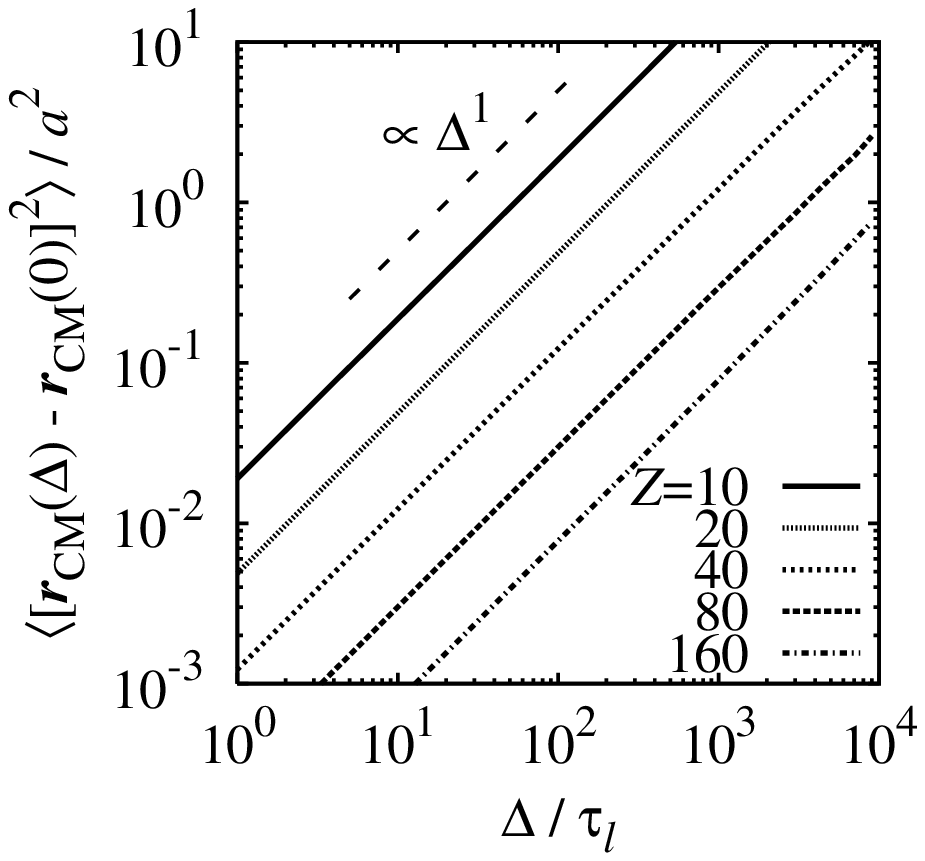}
\caption{\label{discrete_reptation_eamsd}}
\end{figure}

For simplicity, we employ the discrete version of the tube model. (In
the followings we call this model as the discrete reptation model.)
As described in Appendix~\ref{appendix_discrete_reptation_model}, we
model the tube as a freely jointed chain type model. A tube consists of
$Z$ tube segments and each tube segment has a fixed length $a$. (The
number of segments in a tube is fixed to be constant and thus the
contour length fluctuation (CLF)\cite{Doi1986} is absent in this model.)
The polymer chain inside the tube moves (reptates) forth or back
randomly, and we express the characteristic time scale of this random motion
as $\tau_{l} \equiv \zeta_{l} a^{2} /
k_{B} T$ ($\zeta_{l}$ is the friction coefficient of
a segment in the logitudinal direction along a tube).
We use $a$ and $\tau_{l}$ as elementary length and time scales.
The time evolution of a tube is modeled as a random jump type
process. The chain inside the tube reptates forth or back by a given
jump rate, and then a segment at an end is destructed and a new segment
is constructed at another end.
We perform kinetic Monte Carlo simulations for various different values
of $Z$. ($Z$ is proportional to the number of chain segments $N$.) The
number of polymer chains (or tubes) is $1024$ 
for each simulation, and the initial state is sampled from an
equilibrium distribution.
Because this model lacks
the CLF mechanism, the MSD of the CM becomes the
following form\cite{Doi1978}, instead of Eq.~\eqref{tube_model_msd_cm}:
\begin{equation}
 \label{tube_model_msd_cm_without_clf}
 \langle [\bm{r}_{\text{CM}}(t) - \bm{r}_{\text{CM}}(0)]^{2}\rangle
  \propto t.
\end{equation}
The EAMSDs of CMs are shown in Fig.~\ref{discrete_reptation_eamsd}. All EAMSDs
 are proportional to $t$ in all the time regions, which is
consistent with Eq.~\eqref{tube_model_msd_cm_without_clf}.
Therefore, in the discrete reptation model, we cannot obtain any
characteristic time scales from the EAMSD at all.

\begin{figure}[h]
 \includegraphics[width=1.0\linewidth,clip]{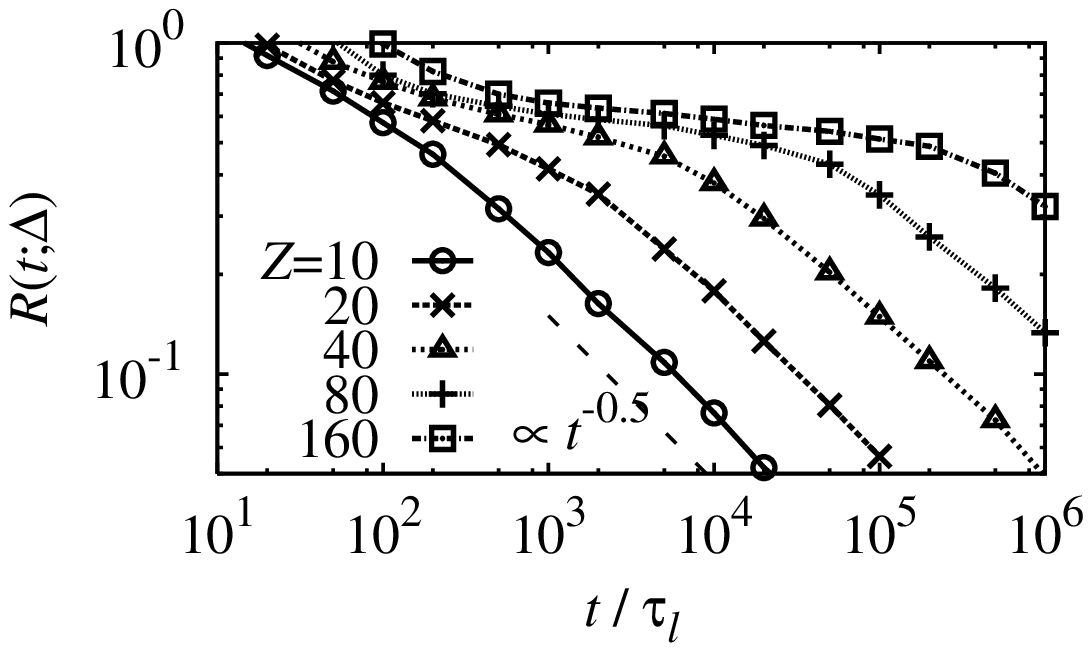}
\caption{\label{discrete_reptation_tamsd_fluctuation_delta10}}
\end{figure}

\begin{figure}[h]
 \includegraphics[width=1.0\linewidth,clip]{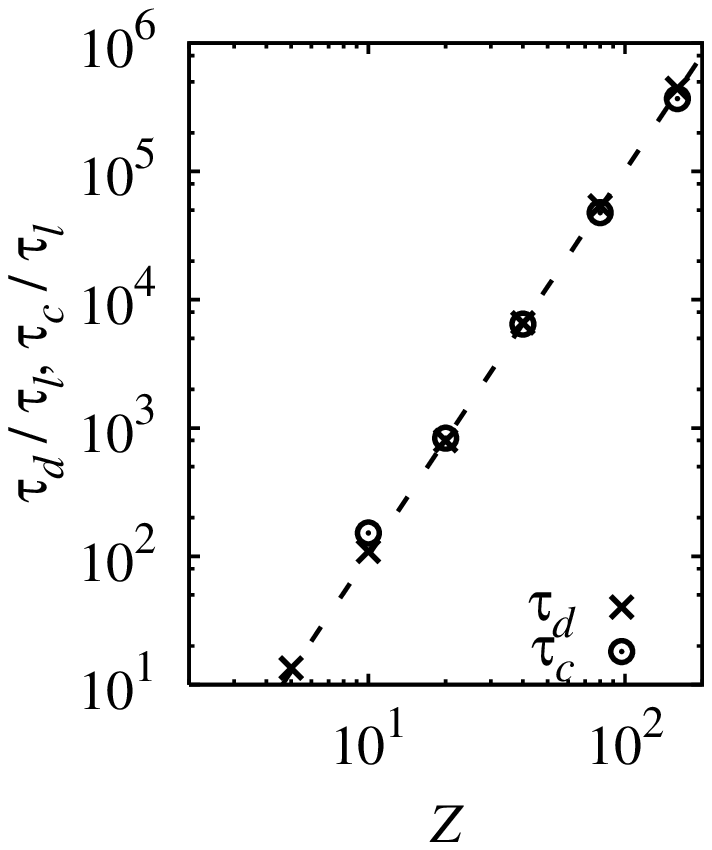}
 \caption{\label{discrete_reptation_relaxation_and_crossover_times}}
\end{figure}

\begin{figure}[h]
 \includegraphics[width=1.0\linewidth,clip]{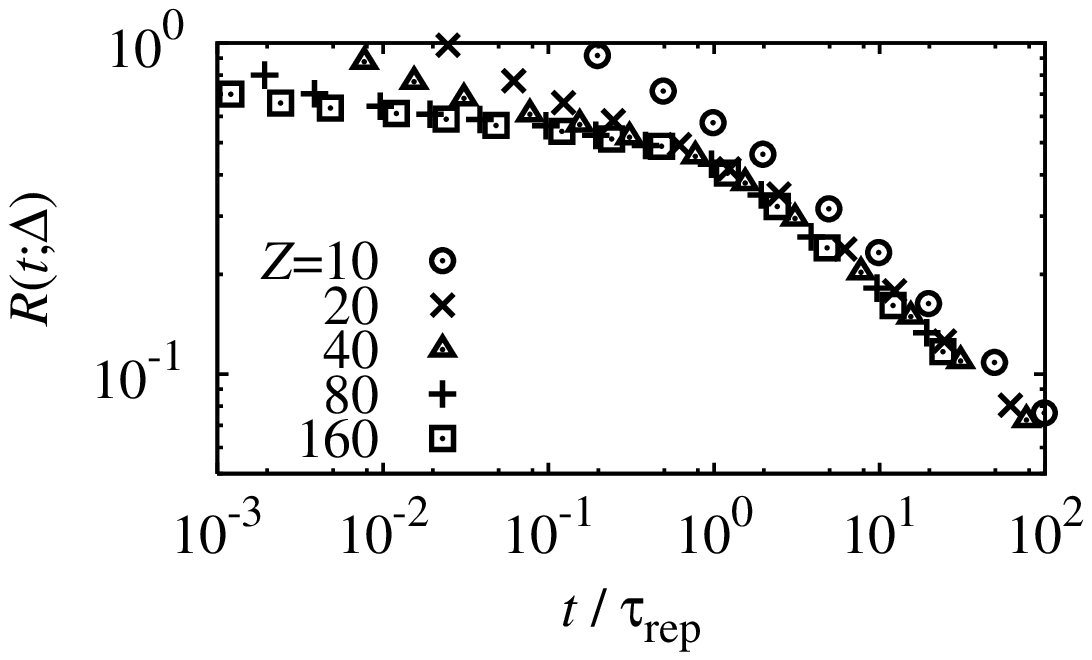}
 \caption{\label{discrete_reptation_tamsd_fluctuation_delta10_rescaled}}
\end{figure}

Interestingly, even in this case, the RF of the TAMSDs shows a clear crossover.
As in the case of the slip-spring model, an appropriate value should be
chosen for $\Delta$ in the RFA. Here we set $\Delta / \tau_{l} = 10$. (We show 
the effect of $\Delta$ value in Appendix \ref{appendix_effect_of_time_interval_size_delta}.)
As shown in Fig.~\ref{discrete_reptation_tamsd_fluctuation_delta10},
the RF is fitted well with Eq.~\eqref{relative_fluctuation_tamsd_cm}. Unlike the case of the slip-spring model, 
the exponent $\alpha$ is almost zero,
{which is different from the results of the slip-spring model.}
Moreover, we observe that the
crossover time strongly depends on $Z$ in a similar way
to the case
of the slip-spring model. The crossover time $\tau_{c}$
obtained from the RF of the TAMSDs are shown in
Fig.~\ref{discrete_reptation_relaxation_and_crossover_times}. The obtained $\tau_{c}$ data
almost coincide with the longest relaxation time $\tau_{d}$ or the
reptation time $\tau_{\text{rep}} = \tau_{l} Z^{3} / \pi^{2}$.
This becomes clearer if we rescale the time $t$ by the reptation time
$\tau_{\text{rep}}$.
Figure~\ref{discrete_reptation_tamsd_fluctuation_delta10_rescaled} shows
the RFs of TAMSDs (the same data as
Figure~\ref{discrete_reptation_tamsd_fluctuation_delta10}) for the
rescaled time $t / \tau_{\text{rep}}$. 
 Interestingly, the rescaled RF data for different $Z$ collapse into one
master curve if $Z$ is sufficiently large, except the short time region.

\section{Discussions}
\label{discussions}


Both the slip-spring model and the discrete reptation model in
Section~\ref{simulation_results} give
qualitatively the same crossover behavior.
{
However, there are some differences between the RFA results of the
slip-spring and discrete reptation models. For example, the silp-spring
model exhibits the power-law type behavior of $R(t;\Delta)$ for $t
\lesssim \tau_{c}$ while the discrete reptation model exhibits almost
constant $R(t;\Delta)$ for $t \lesssim \tau_{c}$. We expect that the
differences reflect the detailed relaxation mechanisms of models.
(In the case of the CTRW, $R(t;\Delta)$ reflects some informations of traps\cite{ctrw-inpreparation}.)

The discrete reptation model is the simple model and the
reptation is only the relaxation mechanism in the model.
On the other
hand, the slip-spring model has other relaxation mechanisms such as the
Rouse motion of subchains and the contour length fluctuation.
The incorporation of additional relaxation mechanisms other than the
reptation modulates several dynamical
behavior\cite{Doi1986,Watanabe1999,McLeish2002,Likhtman2002}.
In the reptation model, the dynamic equation of the CM is described by using
the end-to-end vector\cite{Doi1978}. Thus
the RFs can be also modulated by various relaxation mechanisms,
for example, through the dynamics of the end-to-end vector.
If our expectation is correct, we will be able to extract the
information about the relaxation mechanisms from the RFA. However,
currently we have no analytical theory for the RFs in entangled polymer sysmtes.
Unfortunately, it seems to be difficult to obtain analytical expressions of
RFAs of CMs, and the development of theoretical tools for RFs is desired.}

Although we have examined only two models and some
mechanisms are not considered in them, we believe that we will observe
similar behavior of TAMSDs for other models of entangled
polymers. Judging from our simulation data, the reptation motion is the
most important for the behavior of the RFs. Therefore, even if various mechanisms which affect the
relaxation behavior are incorporated, we expect that the RFA still gives
the information about the long time relaxation due to the reptation.
It should be pointed that several mechanisms such as the CR affect the
longest relaxation time
$\tau_{d}$\cite{Watanabe1999,McLeish2002,Likhtman2002}. We expect that the crossover time $\tau_{c}$ will
be affected in a similar way, and thus the RFA will give qualitatively
the same results for more elaborated models. The numerical factor would
be largely affected by mechanisms such as the CLF and the CR. The
discrepancy between $\tau_{d}$ and $\tau_{c}$ observed in the
slip-spring model may be due to such additional relaxation mechanisms.
From these considerations, we expect that the RFA performed in Section \ref{simulation_results} can be
applied for other models, including the Kremer-Grest model\cite{Kremer1990} or highly
coarse-grained models\cite{Kindt2007,Padding2011}.

{Here we shortly discuss the power-law exponent $\alpha$ of the
slip-spring model (Eq.~\eqref{relative_fluctuation_tamsd_cm}). As we have shown in Section~\ref{slip_spring_model}, it depends on $N$.
As far as the authors know, there is no relaxation mechanism which gives
such $N$-dependent power-law exponent.
Therefore, it would be physically
natural to interpret that this $N$-dependent power-law exponent is apparent. As we can observe in
Fig.~\ref{slip_spring_tamsd_fluctuation_delta10}(b), the RF becomes
proportional to $t^{-0.5}$ for a very small $t$ region. The region where
we have observed the power-law exponent $\alpha$ is the transient
region. If $N$ is relatively small, such a transient region is affected
by the short time region, and the fitting may give apparent power-law
exponent which is different from the true value. (This situation may be
somewhat similar to the power-law for the longest relaxation time,
$\tau_{d} \propto N^{3.4}$, which is considered to be apparent
behavior\cite{Doi1986}.)}

One advantage of the RFA is that
it can be performed only with the CM trajectory data.
As we mentioned, for some cases, the CM trajectory data
are more suitable for the calculation of the long time relaxation
behavior. The accurate computation of the stress autocorrelation
functions in multi chain models is not easy.
In highly coarse-grained models (such as the RaPiD model), the entangled
polymers are modeled with very limited degrees of freedom, and
some information such as the end-to-end vectors
cannot be calculated. The RFA can be applied even for such cases.
Of course, the RFA will not always be the optimal method, and
other methods which utilize conventional quantities will be better for some cases.
The RFA of the CM trajectories should be understood as a complementary
method to conventional ones.
Here we note that in conventional methods to determine the longest relaxation time
$\tau_{d}$, dynamical quantities are utilized after the ensemble
averages are taken. On the other hand, the RFA
of the TAMSDs uses the time-averaged quantity before the ensemble average is
taken. The RFA for other quantities, which are usually analyzed after the
ensemble averages are taken, may be possible.

Another advantage of the RFA will be that we can
utilize it to validate some phenomenological coarse-grained models.
As we will discuss in the next subsection, there are many possible ways
to model the dynamics of entangled polymers. The RF
behavior of TAMSDs will be useful to check the validity of a
phenomenologically constructed dynamical model. That is, if the target
model reasonably reproduces the dynamics of entangled polymers, it
should exhibit the crossover behavior and the crossover time $\tau_{c}$
should be comparable to the longest relaxation time
$\tau_{d}$ (estimated from other physical quantities).
For such a validation, usually the ensemble-averaged
quantities (such as the relaxation modulus) are utilized. We expect that
the RFA, which utilizes a time-averaged quantity, may become a complementary
validation tool.

\section{Conclusions}

In this work we have shown
evidences that the RFA of the TAMSDs of CMs of polymer chains can be
utilized to study the long time relaxation behavior of entangled polymer
systems. The information of the longest relaxation time (the
disengagement time) $\tau_{d}$ is successfully extracted
as the crossover time $\tau_{c}$ in the RF of the CM
trajectories.

The RFA can extract the characteristic long time relaxation behavior of entangled
polymers for two different models (the slip-spring model and the discrete
reptation model), although the behavior somewhat depends on the model
details or parameters. For the discrete reptation model, the
crossover time determined from the RF of the TAMSDs almost
coincide with the longest relaxation time determined from the shear
relaxation modulus.

Although in this work we provided results only for two models, we believe
that our method can be applied for other models of entangled polymers,
including molecular dynamics models and highly coarse-grained
models.
{Further analyses for various simulation data and development of theories
for RFs in entangled polymer systems are demanding.}

\section*{Acknowledgment}
This work is supported by the Core Research for the Evolution Science and
Technology (CREST) of the Japan Science and Technology Agency (JST).

\appendix

\section{Effect of Time {Difference} $\Delta$}
\label{appendix_effect_of_time_interval_size_delta}

\begin{figure}[h]
 \includegraphics[width=1.0\linewidth,clip]{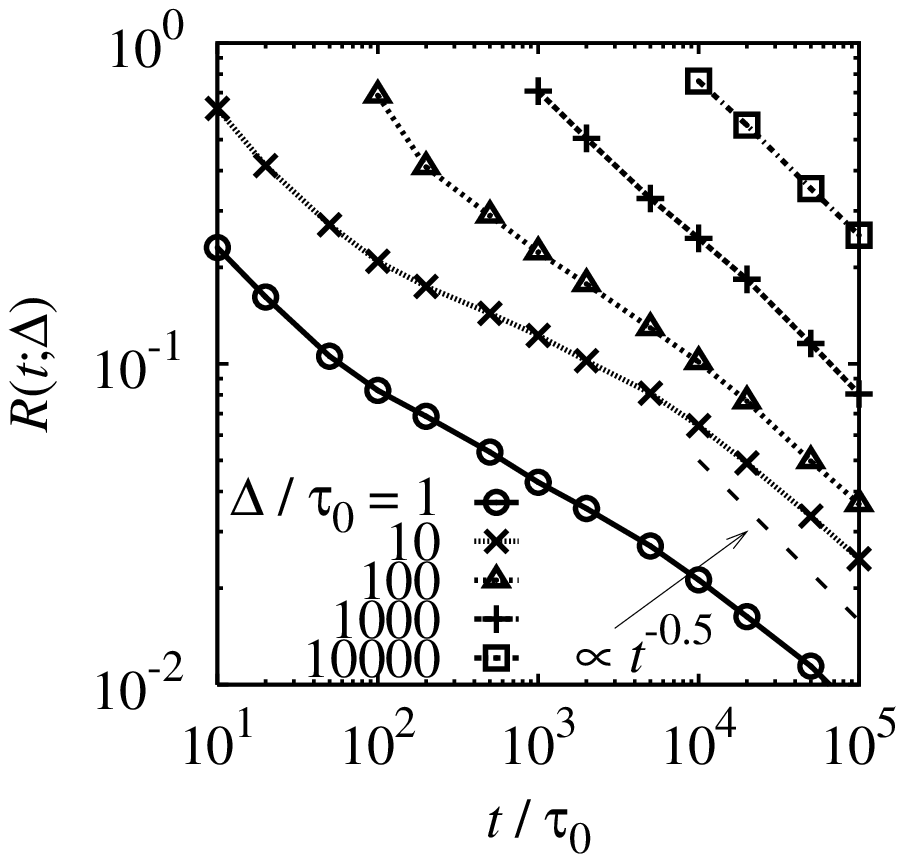}
 \caption{\label{slip_spring_tamsd_fluctuation_n80}}
\end{figure}

In this appendix, we examine the effect of the time
difference size $\Delta$ on the RF of the TAMSDs, $R(t;\Delta)$.
Figure~\ref{slip_spring_tamsd_fluctuation_n80} shows the RF for $N = 80$
and different $\Delta$ values in the slip-spring model.  The forms of
$R(t;\Delta)$ data depend on $\Delta$ rather strongly.
If $\Delta$ is too large (such as $\Delta / \tau_{0} = 1000$), we cannot determine the crossover time
$\tau_{c}$ from the RF data. This may be because the
$\Delta$ value becomes close to $\tau_{d}$ and the crossover behavior
may be smeared out.
If $\Delta$ is too small (such as $\Delta / \tau_{0} = 1$), the
$R(t;\Delta)$ value becomes very small and thus it becomes difficult to
determine $\tau_{c}$ accurately. Thus $\Delta$ is required to be an
intermediate value. In this case, the curve for $\Delta / \tau_{0} = 10$
seems to be better than other curves. For other $N$ cases, we observe a
similar trend. Therefore, in this work we employ $\Delta / \tau_{0} = 10$
to determine the crossover time $\tau_{c}$.  Physically, this value is
comparable to the entanglement time $\tau_{e}$.  (We note that even if we
employ other value for $\Delta$, for example $\Delta / \tau_{0} = 100$, the
result in the main text is qualitatively not affected.)

\begin{figure}[h]
 \includegraphics[width=1.0\linewidth,clip]{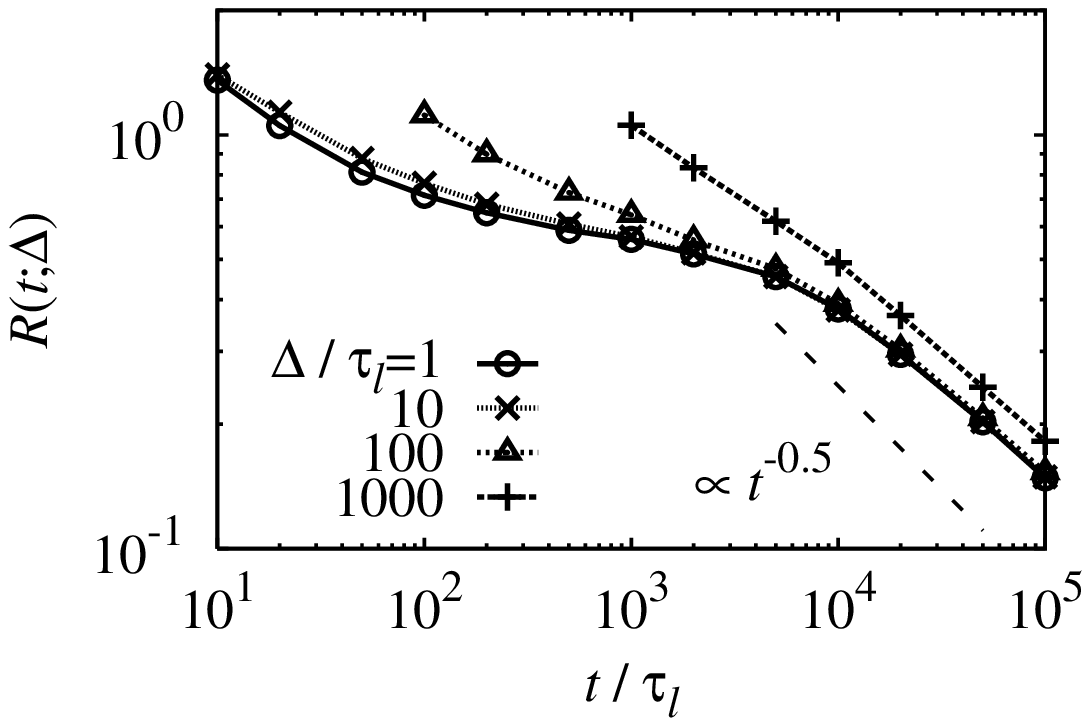}
\caption{ \label{discrete_reptation_tamsd_fluctuation_z40}}
\end{figure}

Figure~\ref{discrete_reptation_tamsd_fluctuation_z40} shows the RF for $Z = 40$ and several
different $\Delta$ values in the discrete reptation model.
As clearly observed in
Fig.~\ref{discrete_reptation_tamsd_fluctuation_z40}, the $R(t;\Delta)$
data for relatively small $\Delta$ ($\Delta / \tau_{l} \lesssim 10$) are
almost the same. Thus we consider that there is a threshold for $\Delta
/ \tau_{l}$ and if $\Delta$ is smaller than the threshold,
$R(t;\Delta)$ is insensitive to $\Delta$. The threshold value depends on several parameters such as $Z$. 
To determine the crossover time $\tau_{c}$ for different $Z$ values,  we should choose $\Delta$
 smaller than the threshold for all the $Z$ values examined.
In this work, we employ $\Delta / \tau_{l} = 10$, which is sufficiently
small for all the examined parameters in the main text. In this case,
$\tau_{l}$ is comparable to $\tau_{e}$ and thus this $\Delta$ is
slightly larger than $\tau_{e}$.

In conclusion, the results in the main text are not so sensitive to the
values of $\Delta$, as long as $\Delta$ is selected to be an
intermediate value.
Results in this appendix implies that the $\Delta$ value should be
selected comparable to (or slightly larger than) the entanglement time $\tau_{e}$. Although the
reason why such $\Delta$ values  work well is not clear, this may be
useful to estimate the optimal value of $\Delta$ when one performs the RFA
for other simulation results. A qualitatively similar
result is obtained by a theoretical analysis for the
CTRW\cite{ctrw-inpreparation}.

\section{Discrete Reptation Model}
\label{appendix_discrete_reptation_model}

In this appendix, we provide the details of the discrete
reptation model and the numerical scheme used for simulations.
In the Doi-Edwards type reptation model\cite{Doi1986}, the polymer chain
is assumed to be confined in a tube which consists of $Z$ discrete
steps. Each step has a constant size $a$, and thus the tube can be
regarded as a freely jointed chain like object\cite{Doi1986}. In this work we express
the positions of tube ends and kinks by $\lbrace \bm{R}_{i} \rbrace$ ($i
= 0,1,2,\dots,Z$). The chain inside the
tube can move only to the direction along the tube (which we call the longitudinal
direction), and this reptation motion is modeled by the Langevin
equation. However, the Langevin equation is not suitable to
describe the dynamics of discrete objects, especially when we perform
numerical simulations. In this work, we employ
the discrete jump dynamics to model the reptation motion.

In the discretized reptation model, the polymer chain can move forth or
back along the chain by the constant segment size $a$ with one
step.  We ignore the contour length fluctuation (CLF) effect and
 the constraint release (CR) effect. Thus
the dynamics of the chain is described only by the reptation motion
described here.  The jump rates are proportional to $Z^{-1}$, because the
jump rate corresponds to the longitudinal motion of a whole chain along the
tube and the effective friction coefficient of a chain is proportional to
$Z^{-1}$. The forth or back jump rates (transition rates) are given as
\begin{equation}
 W_{\pm} = \frac{1}{Z \tau_{l}}
\end{equation}
where $\tau_{l}$ is the characteristic longitudinal diffusion time
defined as $\tau_{l} \equiv \zeta_{l} a^{2} / k_{B} T$ ($\zeta_{l}$ is
the friction coefficient of a segment in the longitudinal direction). Subscripts
``$+$'' and ``$-$'' represents the forth and back directions, respectively.
This transition rate model successfully reproduces the diffusion
coefficient of a chain along the tube (in the
longitudinal direction), $D_{l} = k_{B} T / Z \zeta_{l}$\cite{Doi1978}.

To avoid ambiguities or problems associated with the dynamical mapping
of Monte Carlo simulations\cite{Binder1997,Chatterjee2007}, we employ the kinetic Monte
Carlo scheme\cite{Gillespie1976,Gillespie2007,Chatterjee2007} which can
handle the time evolution of discrete stochastic jump processes exactly.
We start from the equilibrium tube conformation $\lbrace \bm{R}_{i}
\rbrace$ at $t = t_{0}$.
To generate random numbers, we employ the Mersenne twister pseudo
random number generator\cite{Matsumoto1998}.
The kinetic Monte Carlo scheme evolves the
system from time $t = t_{j}$ to $t = t_{j + 1}$ where $j$ is the number
of Monte Carlo steps.
We need the total transition rate $W_{\text{tot}}$ and the (normalized)
probabilities of transitions $P_{\pm}$, to simulate the time evolution. They are given as
\begin{align}
 & \label{total_transition_rate_kmc}
 W_{\text{tot}} = W_{+} + W_{-} = \frac{2}{Z \tau_{l}}, \\
 & \label{normalized_probability_kmc}
 P_{\pm} = \frac{W_{\pm}}{W_{\text{tot}}} = \frac{1}{2}.
\end{align}
The direction (forth or back) is randomly selected according to the
probability $P_{\pm}$. From Eq.~\eqref{normalized_probability_kmc} the
direction is selected with the same probability, and this can be
easily realized by using a uniformly distributed random number.
The time step size is sampled from the exponential distribution. The
time at the $(j + 1)$-th step becomes
\begin{equation}
 t_{j + 1} = t_{j} - \frac{1}{W_{\text{tot}}} \ln u
\end{equation}
where $u$ is the random number sampled from the uniform distribution
in $(0,1)$. The positions of tube ends and kinks at the $(j + 1)$-th
step are given as
\begin{equation}
 \bm{R}_{i}(t_{j + 1}) = 
  \begin{cases}
   \bm{R}_{i + 1}(t_{j}) & (i < Z) \\
   \bm{R}_{Z}(t_{j}) + a \bm{n} & (i = Z)
  \end{cases}
  \quad \text{or} \quad
  \begin{cases}
   \bm{R}_{0}(t_{j}) + a \bm{n} & (i = 0) \\
   \bm{R}_{i - 1}(t_{j}) & (i > 0)
  \end{cases}
\end{equation}
where $\bm{n}$ is the random vector on the three dimensional unit sphere
($|\bm{n}| = 1$). The time series of the tube conformation 
are obtained
by successively iterating the kinetic Monte Carlo time steps.

The segments of the polymer chain are assumed to be uniformly
distributed along the tube. Then the CM position is calculated from
$\lbrace \bm{R}_{i} \rbrace$ as
\begin{equation}
 \bm{r}_{\text{CM}} = \frac{1}{Z} \sum_{i = 0}^{Z - 1} \frac{\bm{R}_{i}
  + \bm{R}_{i + 1}}{2} .
\end{equation}

For the discrete reptation model, we define the
longest relaxation time $\tau_{d}$ by using the shear relaxation modulus
$G(t)$ at the long time limit,
\begin{equation}
 \label{discrete_reptation_longest_relaxation_time_definition}
 - \frac{1}{t} \ln \frac{G(t)}{G(0)} \to \frac{1}{\tau_{d}} \quad (t \to \infty) .
\end{equation}
(This method to determine $\tau_{d}$ is consistent with one
used for the slip-spring model.)
We calculate the shear relaxation modulus by the linear response formula:
\begin{equation}
 \label{linear_response_shear_relaxation_modulus}
 G(t) = c_{0} k_{B} T \langle \sigma_{xy}(t) \sigma_{xy}(0) \rangle .
\end{equation}
Here $c_{0}$ is the spatial average chain number density and
$\sigma_{xy}$ is the $xy$ component of the stress tensor.
The stress tensor $\bm{\sigma}$ is calculated as
\begin{equation}
 \label{discrete_reptation_stress_tensor}
 \bm{\sigma} = \frac{3 k_{B} T}{a^{2}} \sum_{i = 0}^{Z - 1}
  (\bm{R}_{i + 1} - \bm{R}_{i}) (\bm{R}_{i + 1} - \bm{R}_{i}) - (Z + 1)
  k_{B} T \bm{1} .
\end{equation}
We note that
Eqs.~\eqref{discrete_reptation_longest_relaxation_time_definition}-\eqref{discrete_reptation_stress_tensor}
give the shear relaxation modulus which is proportional to the tube
surviving probability\cite{Doi1986}. Therefore it is almost obvious for our model
that the longest relaxation time $\tau_{d}$ coincides to the reptation
time $\tau_{\text{rep}} = \zeta_{l} Z^{3} a^{2}/ \pi^{2} k_{B} T = Z^{3} \tau_{l} / \pi^{2}$.

%

\clearpage

\section*{Figure Captions}

\hspace{-\parindent}%
Figure \ref{slip_spring_eamsd}:
Ensemble-averaged EAMSDs of CMs in the slip-spring model for
 $N = 10, 20, 40, 80,$ and $160$. Dashed lines represent curves which
 are proportional to
 $\Delta^{1}$ and $\Delta^{1/2}$. $b$ and $\tau_{0}$ are the
 size and characteristic time of a segment, respectively.

\

\hspace{-\parindent}%
Figure \ref{slip_spring_tamsd_fluctuation_delta10}:
Relative fluctuations of TAMSDs of CMs in the slip-spring model for
 $\Delta / \tau_{0} = 10$ and $N = 10, 20, 40, 80$ and $160$, where
 $\tau_{0}$ is the characteristic time of a segment. {(a) $N = 10$ and
 $20$, and (b) $N = 40, 80$, and $160$.}
 Dashed line represents a curve proportional to
 $t^{-1/2}$. The power law exponents for the short time regions are
 $\alpha = 0.31, 0.24,$ and $0.19$ for $N = 40, 80$, and $160$, respectively.

\

\begin{figure}[h]
 \includegraphics[width=1.0\linewidth,clip]{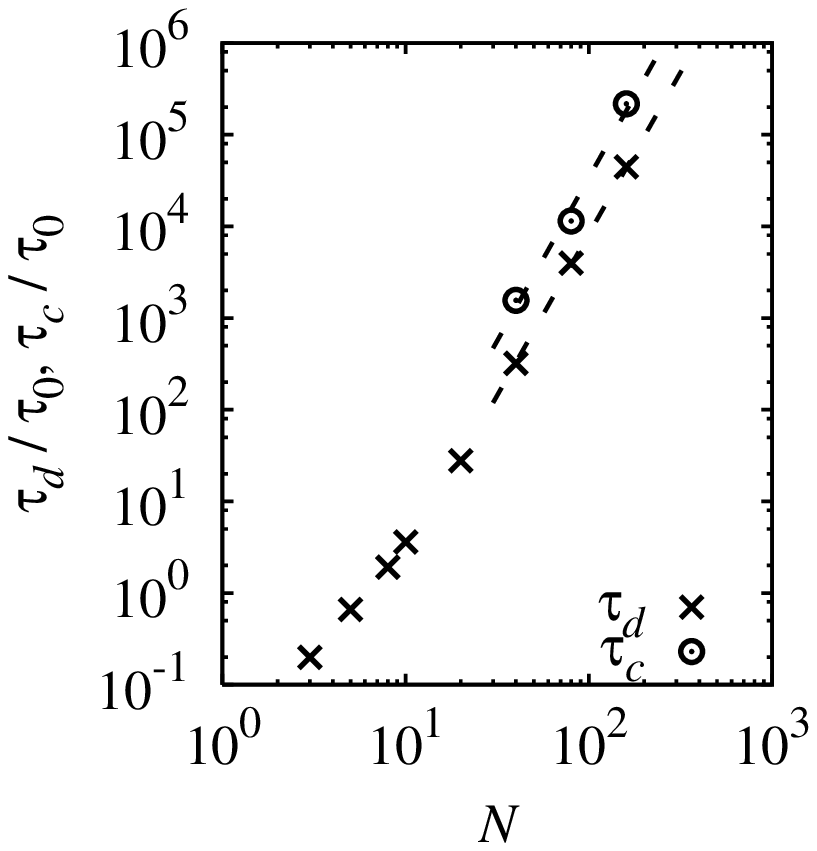}
\caption{\label{slip_spring_relaxation_and_crossover_times}}
\end{figure}

\hspace{-\parindent}%
Figure \ref{slip_spring_relaxation_and_crossover_times}:
The longest relaxation time $\tau_{d}$ and the crossover time
 $\tau_{c}$ in the slip-spring model. $\tau_{d}$ is determined from the
 shear relaxation modulus data \cite{Uneyama2011} whereas $\tau_{c}$ is determined
 from the RF data of TAMSDs for $\Delta / \tau_{0} =
 10$.
 Dashed lines represent fitting curves for large $N$ ($\tau_{d} \propto
 N^{3.48}$ and $\tau_{c} \propto N^{3.51}$).

\

\hspace{-\parindent}%
Figure \ref{discrete_reptation_eamsd}:
Ensemble-averaged MSDs of CMs in the discrete reptation model for
 $Z = 10, 20, 40, 80,$ and $160$. $a$ is the step size of a tube segment
 and $\tau_{l}$ is the characteristic time of the longitudinal motion of
 a segment along the tube.

\

\hspace{-\parindent}%
Figure \ref{discrete_reptation_tamsd_fluctuation_delta10}:
Relative fluctuations of TAMSDs of CMs in the discrete reptation model for
 $\Delta / \tau_{l} = 10$  (see
 Appendix~\ref{appendix_effect_of_time_interval_size_delta}) and $Z =
 10, 20, 40, 80$ and $160$. The dashed line represents a curve proportional to
 $t^{-1/2}$. $\tau_{l}$ is the characteristic time of the longitudinal
 motion of a segment along the tube.

\

\hspace{-\parindent}%
Figure \ref{discrete_reptation_relaxation_and_crossover_times}:
The longest relaxation time $\tau_{d}$ and crossover time
 $\tau_{c}$ in the discrete reptation model. $\tau_{d}$ is determined
 from the shear relaxation modulus and $\tau_{c}$ is determined
 from the RF data of TAMSDs in the same way of Fig.~\ref{slip_spring_relaxation_and_crossover_times}. The dashed line
 represents the reptation time $\tau_{\text{rep}} / \tau_{l} = Z^{3} /
 \pi^{2}$.

\

\hspace{-\parindent}%
Figure \ref{discrete_reptation_tamsd_fluctuation_delta10_rescaled}:
Rescaled RFs of TAMSDs of CMs in the discrete reptation
model for different values of $Z$. The data are the same as
Figure~\ref{discrete_reptation_tamsd_fluctuation_delta10} but the
observation time $t$
is rescaled by the reptation time $\tau_{\text{rep}} = Z^{3} \tau_{l}
/ \pi^{2}$. All the data points collapse into one master curve except the
short time region or the small $Z$ data ($Z = 10$, in this case).

\

\hspace{-\parindent}%
Figure \ref{slip_spring_tamsd_fluctuation_n80}:
Relative fluctuations of TAMSDs of CMs in the slip-spring model for
$N = 80$. The dashed line represents a curve proportional to
$t^{-1/2}$.

\

\hspace{-\parindent}%
Figure  \ref{discrete_reptation_tamsd_fluctuation_z40}:
Relative fluctuations of TAMSDs of CMs in the discrete reptation model for
 $Z = 40$. The dashed line represents a curve proportional to
 $t^{-1/2}$.

\clearpage

\section*{Figures}

\clearpage

\clearpage

\end{document}